\documentclass[a4paper,12pt]{article}

\usepackage{epsfig}
\usepackage{amsmath}
\usepackage{amscd}
\usepackage{amssymb}

\def\bc{\begin{center}}
\def\ec{\end{center}}

\def\be{\begin{equation}}
\def\ee{\end{equation}}
\def\beq{\begin{eqnarray}}
\def\eeq{\end{eqnarray}}
\def\bfig{\begin{figure}}
\def\efig{\end{figure}}
\def\bnum{\begin{enumerate}}
\def\enum{\end{enumerate}}

\begin{document}
\begin{flushright}
Journal-Ref: Astronomy Reports, 2006, v. 50, No 12, pp. 983-1000
\end{flushright}

\vspace{1cm}
\bc
{\Large
\bf An Iterative Method for the Construction of Equilibrium $N$-Body Models for
Stellar Disks} \\
\vspace{0.7cm}
{\bf S.A.~Rodionov (seger@astro.spbu.ru) \\
N.Ya.~Sotnikova (nsot@astro.spbu.ru) }\\
\vspace{0.7cm}
{\it Sobolev Astronomical Institute, St. Petersburg State University,
St. Petersburg}\\
\ec

\vspace{1.0cm}


One widely used technique for the construction of equilibrium models of
stellar disks is based on the Jeans equations and the moments of velocity
distribution functions derived using these equations. Stellar disks
constructed using this technique are shown to be ``not entirely'' in
equilibrium. Our attempt to abandon the epicyclic approximation and the
approximation of infinite isothermal layers, which are commonly adopted in
this technique, failed to improve the situation substantially. We conclude
that the main drawback of techniques based on the Jeans equations is that
the system of equations employed is not closed, and therefore requires
adopting an essentially {\it ad hoc} additional closure condition. A new iterative
approach to constructing equilibrium $N$-body models with a given density
distribution is proposed. The main idea behind this approach is that a model
is first constructed using some approximation method, and is then allowed to
adjust to an equilibrium state with the specified density and 
the required parameters of the velocity distribution
remaining fixed in the process. This iterative approach was used to construct
isotropic, spherically symmetric models and models of stellar disks embedded
in an external potential. The numerical models constructed prove to be close
to equilibrium. It is shown that the commonly adopted assumption that the
profile of the radial velocity dispersion is exponential may be wrong. The
technique proposed can be applied to a wide range of problems involving the
construction of models of stellar systems with various geometries.

\newpage

\section{Introduction}

In studies of the dynamic evolution of galaxies using the
results of numerical $N$-body simulations, it is very important to correctly
specify the initial equilibrium state of the stellar system. Two different
approaches are employed to achieve this. The first approach uses the kinetic
equation (collisionless Boltzmann equation) and the second approach deals with
the
equations of stellar hydrodynamics (Jeans equations). Both approaches have
their advantages and disadvantages.

The kinetic approach is based on the use
of the phase-space distribution function (DF) and the Jeans theorem, which
states that 
any function of integrals of the motion is a solution of the stationary
collisionless Boltzmann equation  \cite{binney-1987}.
For spherically symmetric stellar
systems with isotropic velocity distributions, any function of the form
$f(E)$, where $E$ is the specific energy, describes an equilibrium gravitating
system. If spherically symmetric models with a density distribution profile
$\rho(r)$ that corresponds to the observational data and a potential
$\Phi(r)$ derived 
from the Poisson equation are applied to real objects (elliptical galaxies
and various subsystems of spiral galaxies, such as the bulge and halo),
determining the distribution function $f(E)$
reduces to solving an integral equation derived via the Abel transform
(Eddington's formula; see. e.g., \cite{binney-1987}). Analytical solutions
of this equation 
are known only for special classes of models\footnote{For example, the
distribution function corresponding to the popular 
dark halo model -- the Navarro-Frenk-White \cite{navarro-563-1996} model --
can be determined 
only numerically \cite{widrow-39-2000}, \cite{lokas-155-2001}.}.
Researchers often prefer a
different approach, proceeding from a particular form of $f(E)$ and
integrating it over velocity space to obtain the density distribution and
then the potential. One can then choose from among the broad class of
analytical models constructed in this way those whose density profiles are
closest to the density profile of the real system.

The transition to
anisotropic or axisymmetric models, e.g., stellar-disk models, drastically
complicates the problem. Describing a disk system requires the use of a
distribution function of the form $f(E, L_z)$, where $L_z$ is the angular momentum
of a particle with respect to the symmetry axis $z$. A number of models are
known to have analytical formulas simultaneously for $f(E,L_z)$, $\rho$, and
$\Phi$  (simple
models include the Kalnajs \cite{kalnajs-63-1972} disk, however, this is not
the only one). 
Such models can be used to study the properties of flat systems only as a
first approximation. These are usually 2D models and their radial profiles
usually differ strongly from the exponential form typical of spiral galaxy
disks.

As for 3D axisymmetric models, constructing such models using
the distribution function $f(E, L_z)$ also faces another serious barrier. In
systems represented by such distribution functions, the radial velocity
dispersion should be equal to the velocity dispersion parallel to the
rotation axis (see, e.g., \cite{binney-1987}), which is inconsistent with
observational 
data -- at least with data for our Galaxy based on measurements of stellar
velocity dispersion in the solar neighborhood (see, e.g.,
\cite{dehnen-387-1998}). In the
kinetic approach, axisymmetric models with anisotropic radial and vertical
velocity distributions can be represented by a function of the form
$f(E, L_z, I_3)$, where $I_3$ is the third integral of the motion. The general
expression for $I_3$ is unknown. In sufficiently cool stellar disks, where the
stellar velocity dispersion is small compared to the velocity of the disk's
rotation about the symmetry axis, the energy of vertical oscillations 
$E_z = \Phi(R,z) - \Phi(R,0) + \frac{1}{2}v_z^2$ is a well-conserved quantity
(along almost circular orbits). This can be used as a third integral of
motion when constructing the DF for thin stellar disks with
density distributions $\rho_{\rm disk}(R,z)$, close to the observed exponential
law \cite{kuijken-1341-1995}, \cite{widrow-838-2005}.
Kuijken and Dubinski \cite{kuijken-1341-1995} and Widrow and Dubinski
\cite{widrow-838-2005} also describe a
procedure for correcting the DF in the case of multicomponent
galaxy models. These authors actually solve a more general problem -- the
construction of the distribution function for a disk-galaxy model consisting
of several self-consistently components: an exponential disk,
bulge, dark halo, etc. However, even more specific problems, such as
a construction of an equilibrium model for a disk with a realistic density
profile, which is 
embedded in the external
potential, is not easy to
produce as an initial configuration for $N$-body simulations.

The second approach, which is based on
computing the moments of the equilibrium particle velocity distribution
function, uses the Jeans equations. In the case of constructing equilibrium
stellar-disk models with realistic density profiles, this approach usually
consists of various modifications of the technique described by Hernquist
\cite{hernquist-389-1993}.

This technique has the advantage that it is relatively simple and makes
it possible to construct a model that is more or less close to equilibrium
for a given density distribution profile $\rho_{\rm disk}(R,z)$ and external
potential $\Phi_{\rm ext}(R,z)$, under reasonable assumptions about the
velocity distribution 
function. However, this technique has an important drawback: the resulting
$N$-body model is often far from equilibrium. This primarily concerns models
with small masses for the spheroidal components (bulge and halo). We analyze
the technique of Hernquist in more detail in the next section.

In this
paper, we offer a new iterative method for constructing equilibrium models
of stellar disks embedded in an external (rigid) potential, which we compare
to the widely used method based on the moments of the distribution function.
Our iterative models are very close to equilibrium and have the specified
density profile. The iterative approach has a broader scope than the
particular problem we are solving, and can be used to both construct models
with a different geometry (e.g., spherically symmetric models with a given
mass distribution and anisotropy of the random motions) and construct
equilibrium models of selfconsistent multicomponent stellar systems.

\section{Approach based on the moments of the distribution function}
\label{s_moments}

If a stellar
disk with density $\rho_{\rm disk}(R,z)$ and an external potential
$\Phi_{\rm ext}(R,z)$ produced,
e.g., by the halo and bulge, are axisymmetric, the Jeans equations for the
first and second moments of the velocity distribution function for the disk
particles can be written in the form \cite{binney-1987}, \cite{king-2002}

\be
\label{eq_jeans_1}
\left\{
\begin{array}{rcl}
{\overline v}_{\varphi}^2 &=& v_{\rm c}^2 + \sigma_R^2  - 
\sigma_{\varphi}^2 + \displaystyle \frac{R}
{\rho_{\rm disk}}
\frac{\partial (\rho_{\rm disk} \sigma_R^2)}{\partial R} \, , \\
\sigma_{\varphi}^2 &=& \displaystyle\frac{\sigma_R^2 R}
{2 \overline v_{\varphi}}
\left( \frac{\partial \overline v_{\varphi}}{\partial R} + 
\frac{\overline v_{\varphi}}{R} \right) \, , \\
\displaystyle\frac{\partial (\rho_{\rm disk} \sigma_z^2)}{\partial z} &=& 
-\rho_{\rm disk} 
\displaystyle\frac{\partial \Phi_{\rm tot}}{\partial z} \, ,\\
\end{array}
\right.
\ee
where $\bar v_{\varphi}$ is the mean azimuthal 
velocity\footnote{An overline denotes averaging.}; $\sigma_R$,
$\sigma_{\varphi}$, and $\sigma_z$ are the velocity
dispersions in the radial, azimuthal, and vertical directions,
respectively\footnote{In accordance with the usual astronomical
usage, the dispersion denotes the standard deviations of the distribution
function.};
$\Phi_{\rm tot}=\Phi_{\rm ext} + \Phi_{\rm disk}$ is the total potential 
produced by all components of the system; and 
$v_c = R \displaystyle\frac{\partial \Phi_{\rm tot}}{\partial R}$
is the circular velocity. Here, we have omitted the
dependences of parameters on the coordinates $R$ and $z$ in the cylindrical
coordinate system for simplicity.

Equations (\ref{eq_jeans_1}) assume the absence of
regular motions in the radial and vertical directions. We also assume that
the axes of the velocity ellipsoid are directed along the axes of the
cylindrical coordinate system; this means that second moments of the form
$\overline{v_R v_z}$
are equal to zero. This is a reasonable assumption for the galactic plane
(based on symmetry considerations). However, outside the galactic plane, the
velocity dispersion ellipsoid is inclined \cite{king-2002}, so that the equality 
$\overline{v_R v_z}=0$
breaks down, although substantial deviations appear only in the central
regions of the disk.

The system (\ref{eq_jeans_1}) consists of three equations and has four
unknowns: $\bar v_{\varphi}$, $\sigma_R$, $\sigma_{\varphi}$, and
$\sigma_z$. To solve such a system, one should make some additional 
assumption, and this is the major drawback of methods of constructing
equilibrium models for stellar disks based on the Jeans equations.

One of
the most popular techniques for constructing equilibrium (or, more
precisely, close-to-equilibrium) $N$-body stellar-disk models based on the
Jeans equations is described by Hernquists \cite{hernquist-389-1993}.
Many authors (see, e.g., \cite{revaz-67-2004}, \cite{hoperskov-387-2003},
\cite{athanassoula-35-2002}, \cite{sotnikova-367-2003},
\cite{sotnikova-17-2005}) have applied this technique with various minor
modifications. It is 
usually applied to three-dimensional disks with exponential density profiles:
\be
\label{eq_exp_disk_dens}
\rho_{\rm disk}(R,z)=
\left\{
\begin{array}{ll}
\displaystyle\frac{M_{\rm disk}}{4 \pi h^2 z_0}
\, \exp\left(-\frac{R}{h}\right)
\, {\rm sech}^2\left(
\frac{z}{z_0} \right) &, \: R \le R_{\rm max} \, ,\\
0&, \: R>R_{\rm max} \, .
\end{array}
\right.
\ee
Here, $h$ is the exponential disk scale, $z_0$ is the vertical scale length,
$M_{\rm disk}$ is the total mass of the
disk\footnote{Formally, $M_{\rm disk}$ is the total mass of the disk at
$R_{\rm max}=\infty$.}, and
$R_{\rm max}$ the disk radius
(truncation radius). This density profile approximates well the observed
profiles of real spiral galaxies \cite{kruit-105-1981}.

Further, the conditions under which the Jeans equations (\ref{eq_jeans_1})
were derived should be supplemented with the following
additional assumptions \cite{hernquist-389-1993}.
\bnum
\item
All four moments ($\bar v_{\varphi}$, $\sigma_R$, $\sigma_{\varphi}$, and
$\sigma_z$) are
independent of $z$ and depend only on the cylindrical radius $R$; i.e., the disk
is isothermal in the $z$ direction.
\item 
The epicyclic approximation is valid
(random velocities of stars are small compared to the velocity of rotation).
In this case, the mean azimuthal velocity in the second equation of
(\ref{eq_jeans_1}) can
be replaced by the circular velocity (i.e., we can substitute $v_c$ for
$\bar v_{\varphi}$).
\item
The last equation of (\ref{eq_jeans_1}) is often rewritten in the
approximation of 
infinitesimal isothermal layers \cite{spitzer-325-1942}. In this case, the
contribution of the 
external potential to the total potential $\Phi_{\rm tot}$ is often neglected. This
yields the relation $\sigma_z^2 = \pi G \Sigma_{\rm disk} z_0$, where
$\Sigma_{\rm disk}$  is the
surface density of the disk.
\item
The velocity distribution function has
the Schwartzschild form, i.e., the velocity components along each of the
three axes of the cylindrical coordinate have normal
distributions.
\enum

These are standard assumptions, are believed to be
appropriate for real galaxies \cite{binney-1987}.

The system (\ref{eq_jeans_1}) is closed under the
additional assumptions: $\sigma_R^2 \propto \exp(-R/h)$. 
It is convenient to specify the coefficient
of proportionality 
via the Toomre parameter $Q_{\rm T}$ \cite{toomre-1217-1964} at some radius
$R_{\rm ref}$. This parameter
characterizes the degree of disk heating, or the margin of stability against
the growth of perturbations in the disk plane. For a stellar disk with an
exponential density profile in $R$, the relation 
$\sigma_R^2 \propto \exp(-R/h)$ means that $\sigma_R^2$
is proportional to the surface density $\Sigma_{\rm disk}$. Together with the
isothermal layer approximation, this also yields a dependence of the form
$\sigma_R \propto \sigma_z$. Observational data for our Galaxy
\cite{lewis-139-1989},
\cite{turina-phd} are consistent with the
dependence $\sigma_R^2 \propto \exp(-R/h)$. It is believed
from general considerations that this relations is also appropriate to other
spiral galaxies \cite{bottema-16-1993}. We examine the validity of this
assumption below.

As a result, one can construct for a given $\rho_{\rm disk}$ and $\Phi_{\rm
ext}$ 
a one parametric family of
models parametrized by a quantity that characterizes the degree of disk
heating, or the fraction of the kinetic energy of the disk contained in
random motions.

\begin{figure}
\centerline{\psfig{file=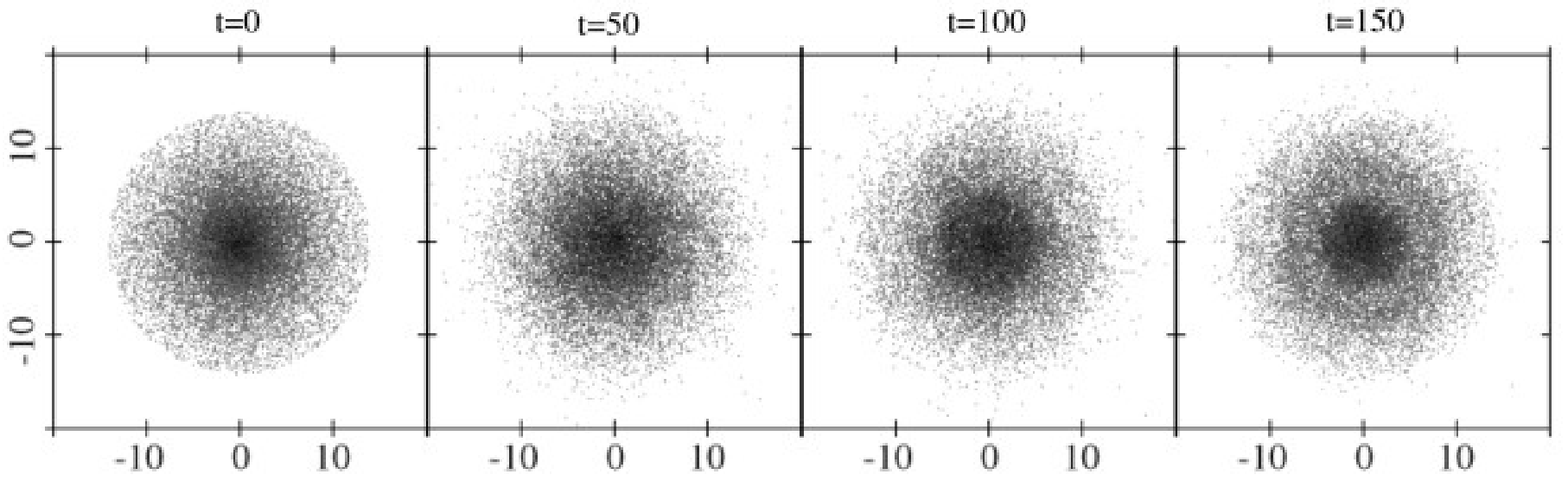,width=16cm}}
\centerline{\psfig{file=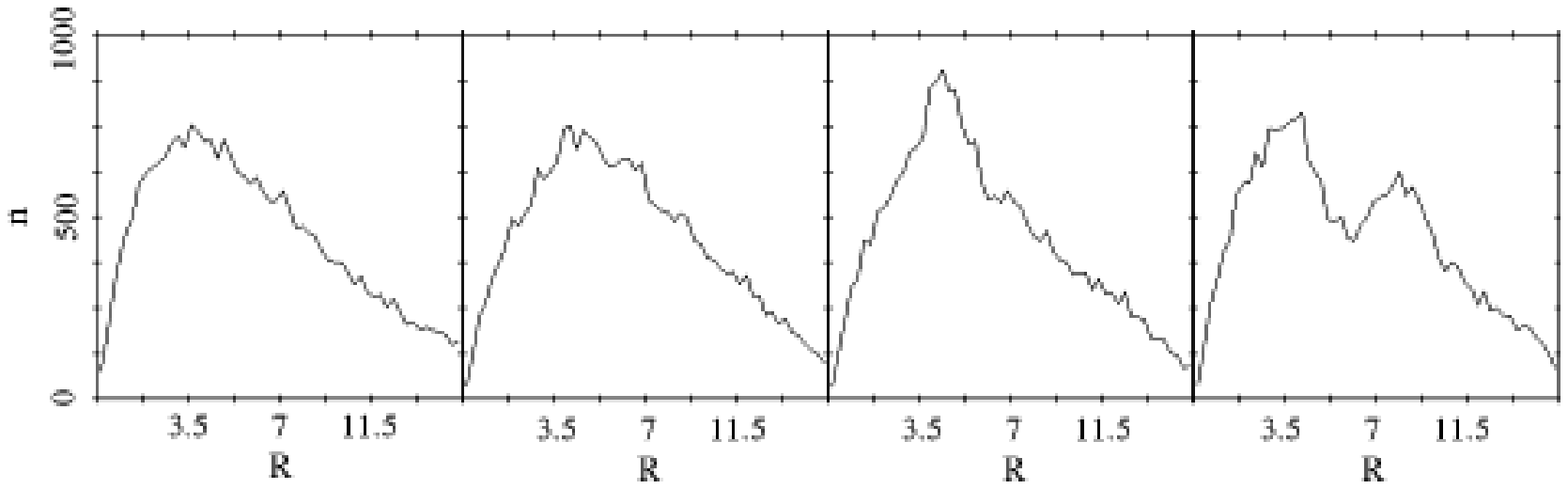,width=16cm}}
\caption{ Adjustment to equilibrium of the model constructed using the
Hernquist technique (see~Section~\ref{s_moments}). The top snapshots show a
face-on view for several times,
with the shades of gray corresponding to the logarithm of the
particle number per pixel.
The plots at the bottom show the radial
distribution of the number of particles (counts made in concentric
cylindrical layers). A characteristic wave propagating from the center can
be seen. The exponential disk model (\ref{eq_exp_disk_dens}) is used with
parameters $h=3.5$, $z_0=1$, $M_{\rm disk}=1$, and $R_{\rm max}=14$.
The external potential has the form of a
Plummer sphere (\ref{eq_plum_phi}) with $a_{\rm pl} = 15$ and 
$M_{\rm pl}=4$. For these parameters, the 
total fractional mass of the spherical component within four exponential
disk scales is about 1.3 
(i.e, $M_{\rm sph}(4h)/M_{\rm disk}(4h) \approx 1.3$). The Toomre
parameter is $Q_{\rm T}(R_{\rm ref})=1.5$, where $R_{\rm ref}=8.5$. 
150 time units
approximately corresponds to one disk rotation period at $R=8.5$. The number
of particles, smoothing parameter, and integration step are 
$N=25000$, $\epsilon=0.02$, and $dt=0.01$, respectively.}
\label{fig_old_model}
\end{figure}

The deficiency of this approach is that the constructed
stellar disks prove to be not entirely in equilibrium. Although the models
rapidly adjust to equilibrium, this adjustment may complicate analyses of
the results of $N$-body simulations, such as those used to study instabilities
of the stellar disk. In the process of this adjustment, a characteristic
ringlike density wave forms and propagates from the center, as is
illustrated by the results of our numerical 
simulations\footnote{In all numerical simulations that used stellar-disk
model (2), we set $G=1$, $h = 3.5$, and $M_{\rm disk}=1$.
If needed, the results of our simulations can
be interpreted in a dimensional system of units (one of several possible) in
which the unit of length is $R_u=1$ kpc and the unit of mass is 
$M_u=8 \cdot 10^{10} M_{\odot}$.
The units of time and velocity are then $t_u \approx 1.67$ Myr 
and $v_u \approx 587$ km$/$s.}
in~Fig.~\ref{fig_old_model} (see \cite{sotnikova-367-2003},
\cite{sotnikova-17-2005} for a more detailed description of the technique
used for the numerical 
simulations). Kuijken and Dubinski \cite{kuijken-1341-1995} also pointed out
a similar effect. 
The most nonequilibrium models are hot models (with large 
$Q_{\rm T}(R_{\rm ref}) \approx 2$) and
models without halos or with low mass halos. The only nearly equilibrium
models are those with fairly massive halos (such as the models described by
Hernquist \cite{hernquist-389-1993}, where the mass of the halo within four
exponential disk 
scales was greater than five disk masses).

The nonequilibrium nature of the
models constructed using the Hernquist technique is due to the underlying
assumptions of this technique. Our
attempt to refine this approach by abandoning the epicyclic and
isothermal layer approximations did not result in any substantial
improvement of the models.

\begin{figure}
\centerline{\psfig{file=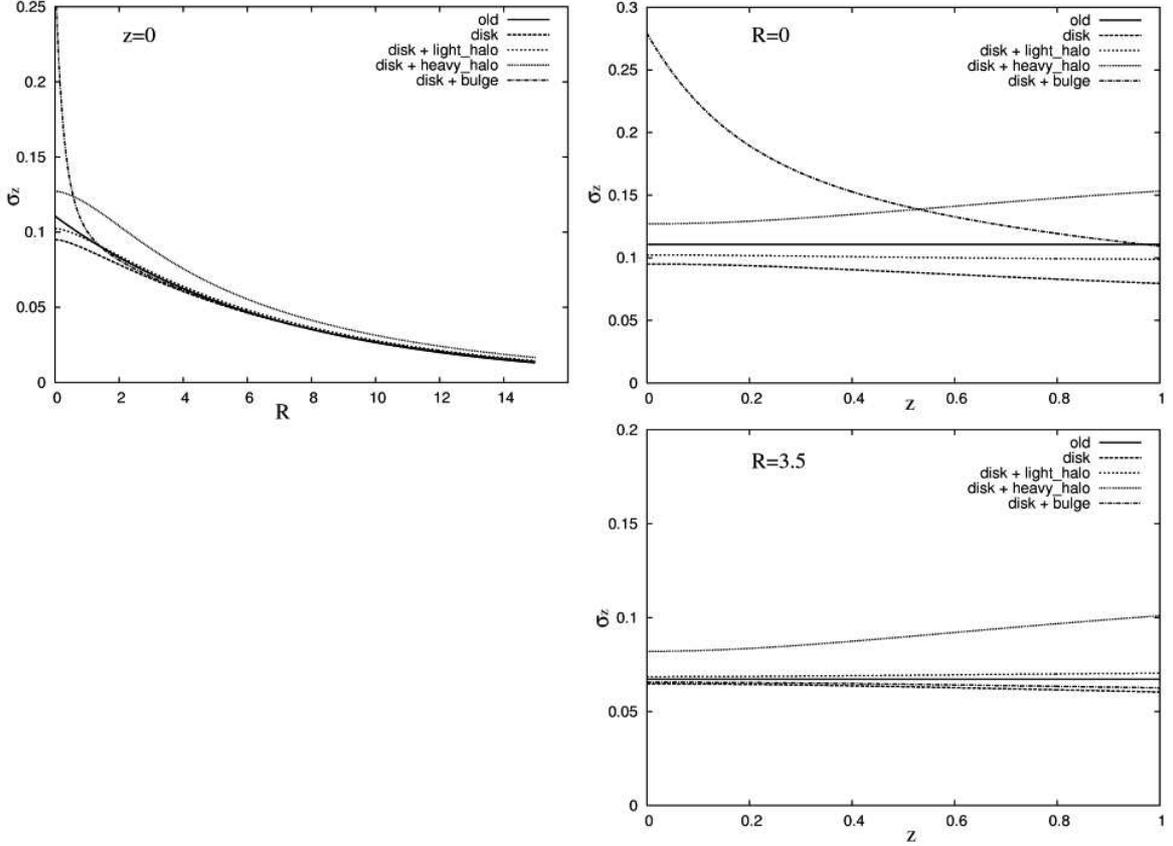,width=16cm}}
\caption{
Dispersion of the vertical stellar velocities, $\sigma_z$, computed both
in the approximation of infinite isothermal layers and without this
approximation for various external potentials. The top left plot shows the
computed dispersion in the disk plane ($z = 0$), the top right plot the
dependence of the dispersion on $z$ at $R = 0$, and the bottom right plot the
dependence of the dispersion on $z$ at $R = 3.5$. In this case, we used a model
of a relatively thin exponential disk (\ref{eq_exp_disk_dens}) 
with $h=3.5$, $z_0=0.3$, $M_{\rm disk}=1$, and $R_{\rm max}=\infty$.
The ``old'' curve shows the dispersion computed in the approximation of
infinite isothermal layers. All other curves were constructed without this
assumption and the corresponding models differ in the adopted external
potentials: the ``disk'' model was computed with no external potential, the
``disk + light halo'' model with an external potential in the form of
Plummer
sphere (\ref{eq_plum_phi}) with $a_{\rm pl}=3.5$ and $M_{\rm pl}=1$, 
the ``disk + heavy halo'' curve with an
external potential in the form of the Plummer sphere with 
$a_{\rm pl}=3.5$ and $M_{\rm pl}=5$,
and the ``disk + bulge'' curve with an external potential in the form of
a Hernquist (\ref{eq_hernq_phi}) sphere with $a_{\rm hr}=0.5$ and $M_{\rm
hr}=0.1$.}
\label{fig_svz_thin}
\end{figure}

For example, the solution of the equation of
vertical equilibrium (the third equation of (\ref{eq_jeans_1}))
and of the equation derived
from it in the isothermal layer approximation yield similar $\sigma_z(R,z)$
in many 
cases, and have little effect on the extent to which the models are
nonequilibrium. The results of the ensuing computations illustrate this
conclusion. Since the equation of vertical equilibrium contains only one
unknown, $\sigma_z$, when $\rho(R,z)$ is specified, this equation can be solved
independently of the other equations\footnote{For example, Revaz and
Pfenniger \cite{revaz-67-2004}, Khoperskov et al. \cite{hoperskov-387-2003},
and Bahcall \cite{bahcall-156-1984} used the solution of this equation in
analyses of the equilibrium in 
the vertical direction and the construction of equilibrium models for
stellar disks. Note that this equation can be solved in two ways. As we did
here, one can specify the density profile and find $\sigma_z(R,z)$.
Alternatively, $\sigma_z(R,z)$ can be specified -- for example, by assuming
that $\sigma_z$ is independent of $z$ \cite{bahcall-156-1984} -- in order to
determine the vertical density profile. Khoperskov et 
al. \cite{hoperskov-387-2003} implemented the latter approach.}.
As is evident
from Fig.~\ref{fig_svz_thin}, the approximation of infinite isothermal
layers breaks down only if the external potential in the system is
represented by a very massive dark halo (the ``disk + heavy halo'' curves) or
varies strongly over a vertical disk scale height, e.g., when the system has
a bulge (the ``disk + bulge'' curves). If the system has no external potential
at all (the ``disk'' curves) or is represented by a small halo with a mass
comparable to the mass of the disk (the ``disk + light halo'' curves), then
the two methods yield virtually the same dispersion (except for the
centermost regions, where the approximation of infinite isothermal layers
obviously breaks down, although, even there, the differences are not very
large).

We believe that the nonequilibrium state of the models is due
primarily to the adopted additional assumption concerning the dependence of
$\sigma_{R}$ on $R$.

Note that a number of authors have adopted other assumptions,
different from the proportionality $\sigma_R^2 \propto \exp(-R/h)$,
or equivalently $\sigma_{R} \propto \sigma_z$, as
the missing equation in the system of Jeans equations. 
For example, Athanassoula and Misiriotis \cite{athanassoula-35-2002} adopted
the additional assumption that the Toomre parameter $Q_{\rm T}$ is
independent of radius.  
Revaz and Pfenniger \cite{revaz-67-2004}
assumed that $\sigma_{R} \propto \sigma_z \displaystyle \frac{\nu}{\kappa}$,
where $\kappa$ is the epicyclic
frequency and $\nu$ is the frequency of the vertical oscillations,
($\nu^2 = \displaystyle\frac{\partial^2 \Phi_{\rm tot}}{\partial z^2}$).
They computed the coefficient of proportionality as
in the original technique of Hernquist, by specifying
the Toomre parameter at some radius.

So far, we have no grounds to prefer
any particular additional condition to supplement the Jeans equations.
However, by taking into account all the moments up to the sixth inclusive,
and if that all the moments beginning with the second are small compared to
the circular velocity, this problem can be solved without invoking
additional assumptions \cite{amendt-79-1991},
\cite{cuddeford-427-1991}. Unfortunately, due to its complexity and
awkwardness, this solution has never been used to construct $N$-body models of
stellar disks.

\section{Dependence of the radial velocity dispersion on the
cylindrical radius (dependence of $\sigma_R$ on $R$)}
\label{s_sigma_R_z}

The assumption that $\sigma_R \propto \sigma_z$ is now
generally accepted, or, in any case, it is used to
analyze and interpret the observation data \cite{bottema-16-1993},
\cite{kregel-481-2005}, \cite{gerssen-545-2000}. However, is this assumption
justified? 

Thus far, the observed dependence of $\sigma_R$ on $R$ has been obtained
only for the Milky Way. Based on a derivation of this dependence from
observations of K giants, Lewis and Freeman \cite{lewis-139-1989} 
concluded that $\sigma_R^2 \propto \exp(-R/h)$. 
In the isothermal layer approximation, this means that 
$\sigma_R \propto \sigma_z$.
However, we must emphasize an important point. The errors in the dispersion
obtained by Lewis and Freeman \cite{lewis-139-1989} 
are fairly large (on the order of 10\%),
as are the errors in the stellar distances, which were derived indirectly.
Accurate (errors of the order of one percent) measurements of $\sigma_R$ are
available only in the solar neighborhood, based on HIPPARCOS data
\cite{dehnen-387-1998}.
Therefore, for the Galaxy, we can only conclude that $\sigma_R$ is
approximately proportional to $\sigma_z$. It seems possible that, if the
dependence of $\sigma_{\varphi}$ on $R$ for
the Galaxy could be measured with the same accuracy as we now know the
dependence for $\sigma_R$, it would also fit the relation
$\sigma_{\varphi} \propto \exp(-R/2h) \propto \sigma_z$. This
situation should improve when new astrometric satellites (in particular,
GAIA) are launched. At least for the Galaxy, the dependence of the moments
of the velocity distribution function on $R$ will then be known more
accurately.

It is not possible to obtain $\sigma_R(R)$ for other galaxies directly
from observations, which can yield only the line-of-sight velocity, 
$v_{\rm los}$,
and line-of-sight velocity dispersion, $\sigma_{\rm los}$. For edge-on
galaxies, $\sigma_{\rm los}$
depends on both $\sigma_R$ and $\sigma_{\varphi}$, whereas, for
intermediate-inclination galaxies, all 
three components of random velocities ($\sigma_R$,  $\sigma_{\varphi}$,
$\sigma_z$) contribute to $\sigma_{\rm los}$. However,
we can use the Jeans equations (\ref{eq_jeans_1}) to derive the unknown
dependence $\sigma_R$ on $R$
from the observed quantities $v_{\rm los}$ and $\sigma_{\rm los}$.
Gerssen et al. \cite{gerssen-618-1997}, \cite{gerssen-545-2000}, 
Shapiro et al. \cite{shapiro-2707-2003}, 
and Wesfallet al. \cite{astro-ph-0508552} used such considerations to infer the
velocity dispersions in three directions from observational data for
inclined galaxies. The drawback of the technique they used is that it
involves an a priori assumption about the form of the dependence of
$\sigma_R$ and $\sigma_z$
on $R$:
\beq
\nonumber
\sigma_z(R) &=& \sigma_{z,0} \, \exp(-R/a) \, ,\\
\sigma_R(R) &=& \sigma_{R,0} \, \exp(-R/a) \, ,
\eeq
where $a$, $\sigma_{z,0}$, and $\sigma_{R,0}$, are parameters determined
during the reduction of the data. These authors pointed out that there are
no grounds to believe that the scale parameter $a$ should be the same for 
$\sigma_z$ and $\sigma_R$, but the quality of the observational data prevents independent
determination of these two scale parameters.

It follows that the quality of
the observational data available so far prevents the determination of the
dependence of $\sigma_R$ on $R$ for external galaxies, and available
observations can 
only yield estimates of the velocity dispersion for the entire galaxy.

We also note the two semi-theoretical papers \cite{bienayme-781-1997} and
\cite{fux-21-1994}, whose authors used
independent methods to conclude that $\sigma_R \propto \sigma_z$ is not
valid for our Galaxy. 

\section{Iterative method for constructing equilibrium models}
\label{s_iterative_method}

 $N$-body simulations
sometimes use the following method to specify the initial conditions. The
initial conditions are specified (in some way) to be close to equilibrium.
The system is then given some time to adjust to a new equilibrium state,
which is used as the initial state for the $N$-body simulations (Sellwood and
Athanassoula \cite{sellwood-195-1986} and Barnes \cite{barnes-699-1988} used
this approach with some 
modifications). The drawback of this technique is that it is difficult to
build a close-to-equilibrium model with a given density profile.

We developed
a method that is essentially a logical extension of the above technique. The
main idea of our method is to allow the system to adjust to the equilibrium
state while fixing the density profile. We achieve this via the following
general algorithm for the iterative method.

\begin{enumerate}
\item
Use some approximate method
to construct a close-to-equilibrium $N$-body model with a given density
distribution (i.e., with a given particle distribution function in 
space)\footnote{This initial model does not have to be close to
equilibrium. In the next
section, we use a cold model (with zero velocities) as an initial
approximation to construct an equilibrium Plummer sphere.}.
\item
Allow the model to evolve for a short time.
\item
Construct a model with
the same velocity distribution as that of the evolved model, but with the
required density profile (the initial density profile). Note that, if we
have certain constraints on the particle velocity distribution (e.g., if we
want to construct a Plummer sphere model with an isotropic velocity
distribution), we must correct the particle velocity distribution to satisfy
these constraints.
\item
Return to item 2. Stop iterations when the velocity
distribution ceases to change.
\end{enumerate}
As a result, we obtain a close-to-equilibrium $N$-body model with the given density profile.

Practical implementation of the
iterative method is somewhat more difficult. The main problem arises at the
third stage -- the construction of a model with the same velocity
distribution as for the slightly evolved model of the previous iteration
step. In the ideal case, we would have to obtain the particle velocity
distribution at each point of the system. This is, naturally, impossible,
given the available number of particles in the $N$-body simulation. However,
this problem can be resolved if we make some simplifying assumptions.

The
next section describes how an iterative method can be used to construct a
spherically symmetric, equilibrium, isotropic model with a given
distribution function (a Plummer sphere). In the problem at hand, this
enables us to verify that the iterative method indeed permits the
construction of an equilibrium model. This realization of the technique can
be used to construct spherically symmetric, isotropic-velocity models with
other density profiles and, with small modifications, even models with
anisotropic velocity distributions. Below we apply the iterative method to
construct equilibrium $N$-body models for stellar disks. The realization of the
third part of the algorithm in the case of spherically symmetric models
differs from that for disk models, but the main idea remains the same.

\subsection{Equilibrium, isotropic, spherically symmetric models}
To test the idea of the
iterative method, we will use it to construct an equilibrium model for a
case when the equilibrium distribution function is known a priori: an
equilibrium, isotropic model of a Plummer sphere. The isotropy means that
there is no special direction in velocity space, i.e., the velocity
distribution function depends only on the speed. The equilibrium
distribution function for this model is known 
(see, e.g., \cite{binney-1987}, p. 223), and
the potential for this model has the form
\be
\label{eq_plum_phi}
\Phi_{\rm pl}(r) = - \frac{G M_{\rm pl}}
{\left(r^2 + a_{\rm pl}^2\right)^{1/2}} \, ,
\ee
where $M_{\rm pl}$ is the total mass for the model and $a_{\rm pl}$ is the
scale length (the 
Plummer model has about $35\%$ of its mass inside the radius $a_{\rm pl}$).

We now
implement the third part of the algorithm of the iterative method (the
``transfer'' of the velocity distribution function) as follows. We take our
slightly evolved model, from which we plan to copy the velocity distribution
function. We subdivide this model into spherical layers containing
approximately the same number of particles and construct the distribution of
the particle speeds $v$ in each of these spherical layers.

We do this in the
usual way, by determining the range of variation of $v$ and subdividing this
interval into some number of bins. We then compute the number of particles
falling in each bin. In the limit of an infinite number of particles and
infinite number of bins, the resulting histogram gives the distribution
function (not normalized to unity). We assumed that the number of particles
in a bin is equal to the value of distribution function at the center of the
bin. We 
can then approximate these values by some function and use this as the
distribution function. We used several types of approximation:
piecewise-linear, cubic spline, and least-squares fits using functions of
the form $\exp(P(x))$, where $P(x)$ is a polynomial. The type
of fit had virtually no effect on the result of the iterative algorithm.

This yields the distribution function of velocity modulus
in each of the spherical layers.
We then constructed the model for the next iteration step as follows. We
specified the positions of the particles in accordance with the given
density profile. We then determined the spherical layer to which each
particle belongs, and used the ``selection-rejection'' method with the
distribution function for the given spherical layer to determine the speed,
choosing the direction for the velocity as random.

The velocity distribution
function for the new model is isotropic and spherically symmetric (here, we
mean that the velocity distribution function is symmetric in coordinate
space\footnote{Isotropy can be considered to be spherical symmetry in
velocity space.}). We did not simply transfer the velocity distribution
function, but adjusted it to make it strictly isotropic and spherically symmetric.

We now
apply the algorithm described above to construct an equilibrium model for
the Plummer sphere. We use virial units ($G = 1$, the total mass of the model
is~$M_{\rm pl} = 1$, and the total energy is~$E = -1/4$). In iterative models, we
truncated the density distribution at $r_{\rm max} = 5$ 
(the sphere of this radius contains about
98\% of the mass of the modeled system).

Let us consider two models constructed using iterative method:

\begin{itemize}
\item
model ``li'' with a piecewise-linear
approximation for the distribution function;
\item
model ``epol,'' in which the
distribution function is approximated by functions of the form $\exp(P(x))$,
where $P(x)$ is a polynomial (we used a sixth-degree polynomial).
\end{itemize}

We constructed both models in 120 iterations, with the first 100 having
relatively low accuracy. The number of particles is $N=10^5$, the number of 
spherical layers used to subdivide the system $n_r=50$, and the number of
bins used to construct the distribution function of the speed $v$ in each 
spherical layers, $n_v=21$. We performed the last 20 iterations with
relatively high accuracy: $N=5 \cdot 10^5$, $n_r=200$, $n_v=31$. 
The time step for
each iteration is $t_i = 1$ (which approximately corresponds to the crossing
time of the system's core in the adopted units). Our starting point was a
cold model with zero velocities.

\begin{figure}
\centerline{\psfig{file=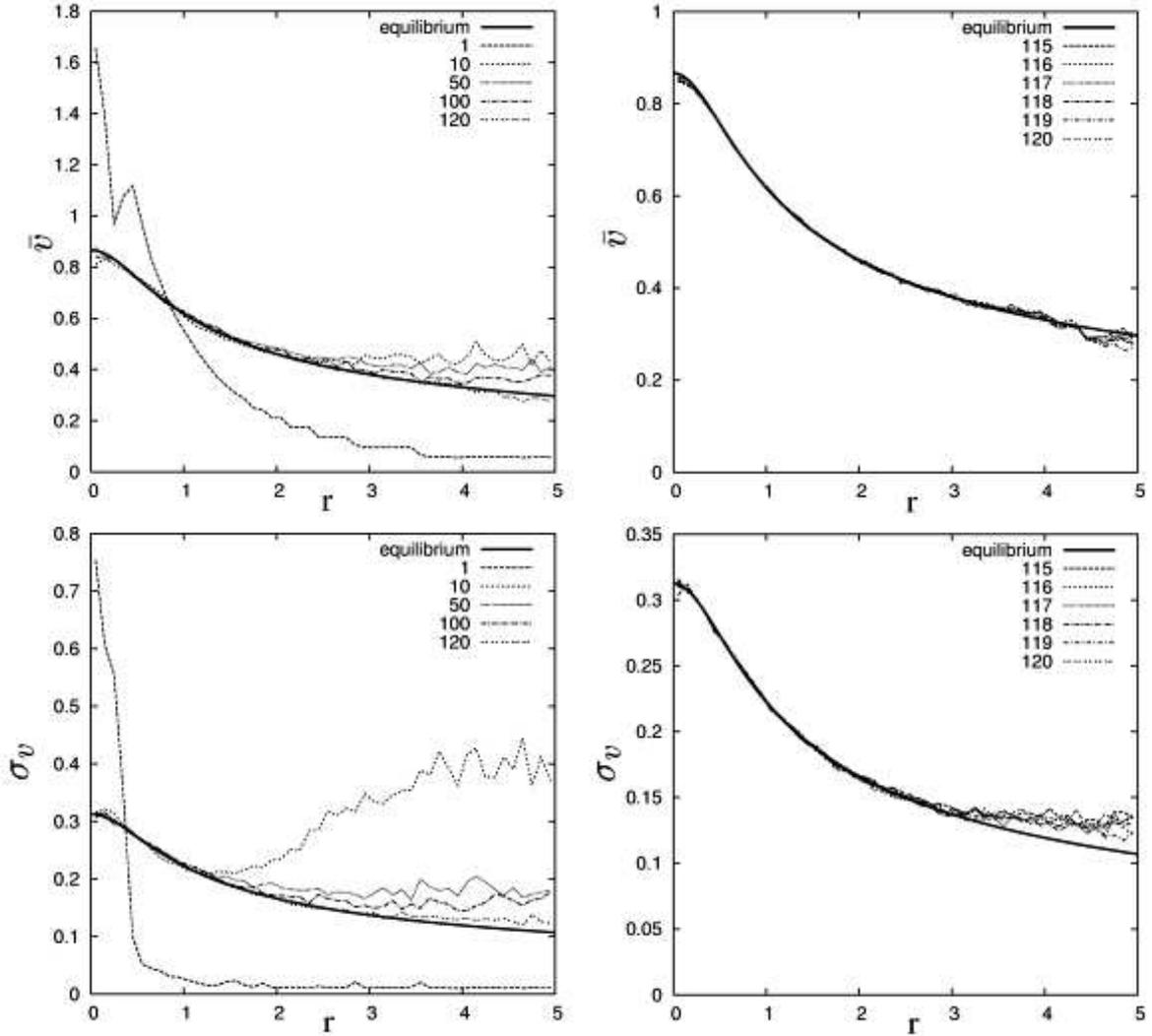,width=16cm}}
\caption{Convergence of iterations in the construction of model li.
The upper
plots show the dependence of the mean speed, $\bar v$, on $r$,
while the lower plots
show the dependence of the dispersion of the speed, $\sigma_v$, 
on $r$ for several
iterations. The left-hand plots show these quantities for the 1st, 10th,
50th, 100th, and 120th iterations, while the right-hand plots correspond to
the last five iterations. The solid bold curve shows $\bar v$ 
and $\sigma_v$ for
equilibrium model (\ref{eq_plum_moments})
(these quantities can be computed by integrating the
equilibrium distribution function).}
\label{fig_plum_conv}
\end{figure}

Note that the initial model in our
computations is far from equilibrium; however, the iterations converge to
close-to-equilibrium models. Figure~\ref{fig_plum_conv}
demonstrates the convergence of the
iterations during the construction of model li. It is clear that the
iterations converge to the equilibrium model. Departures from equilibrium
are evident only at the periphery of the system. For example,
the velocity dispersion, $\sigma_v$, for model li differs appreciably from the value
given by the equilibrium model at $r > 3$ (Fig.~\ref{fig_plum_conv}).
However, the system has
only about 5\% of its mass located outside the sphere of radius $r = 3$.

Let us
now experimentally test the equilibrium of our models by comparing them with
two other models. The first one is constructed using the equilibrium
distribution function as an example of the 
closest-to-equilibrium model (model ``DF''), and the second is
constructed using
another approximation method, based on the following main idea. We can
compute the mean speeds, $\bar v$,
and velocity dispersion, $\sigma_v$, at each radius of
the equilibrium model for the Plummer sphere:
\beq
\nonumber
\bar v &\approx&  0.665 \sqrt{-\Phi_{\rm pl}(r)} \, ,\\
&&
\label{eq_plum_moments}
\\
\nonumber
\sigma_v  &\approx &0.240 \sqrt{-\Phi_{\rm pl}(r)} \, .
\eeq
The numerical coefficients in (\ref{eq_plum_moments}) are determined in
virial units by
integrating the equilibrium distribution function 
(in virial units, $M_{\rm pl} = 1$ and $a_{\rm pl} = 3 \pi / 16$).

An approximate model can be constructed assuming that the
speed distribution function is Gaussian with the moments
(\ref{eq_plum_moments}); we will call
this as the ``gauss'' model.

Let us compare the initial stages of the
evolution of the four $N$-body models for the Plummer sphere (li, epol, DF,
and gauss) for $N=50\,000$. We chose the integration step and smoothing
parameter for the potential $\epsilon$ in
$N$-body simulations in accordance with the
recommendations derived in \cite{rodionov-1-2005} 
($dt=0.002$, $\epsilon = 0.007$).

We analyzed the
results using the $\Delta_r$ and $\Delta_v$ values introduced and determined
as follows (we used similar quantities in our earlier paper
\cite{rodionov-1-2005}). 

The quantity $\Delta_r$
characterizes the spatial deviation of the particles at time $t$ from the
initial distribution (at time $t = 0$). We computed this by subdividing the
model into spherical layers, computing in each spherical layer the
difference between the numbers of particles at time $t$ and at the initial
time, and then computing $\Delta_r$ as the sum of the absolute values of these
differences, normalized to the total number $N$ of particles in the system.
The thickness of the layer and maximum radius were $0.1$ and $2$, respectively.

Note the very important point that the value of $\Delta_r$ for two random
realizations of any model is not zero. This quantity has appreciable natural
noise due to the finite number of particles. We estimated this noise level
by computing the average value of $\Delta_r$ for a large number of pairs (200) of
random realizations of the model at the initial time.

We computed $\Delta_v$ in the
same way, but for the distribution of particles in velocity space. For the
results reported here, the width of the layer and maximum velocity were $0.1$
and $1.5$, respectively.

\begin{figure}
\centerline{\psfig{file=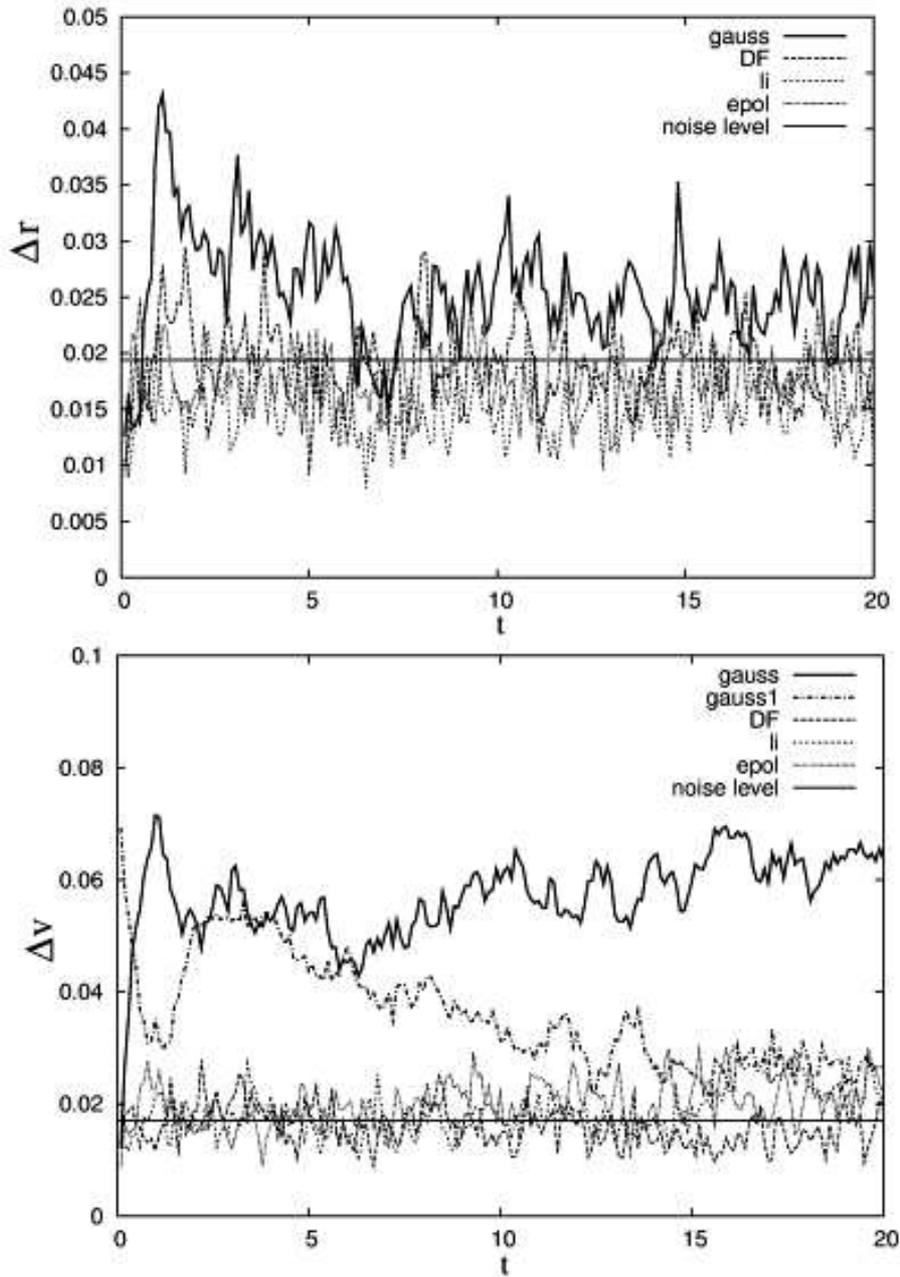,width=12cm}}
\caption{Time dependences of $\Delta_r$ and $\Delta_v$ for four $N$-body 
models of the Plummer
sphere. The horizontal line shows the noise level for $\Delta_r$ and 
$\Delta_v$. The initial
distribution for the gauss model is far from equilibrium (e.g., from the
velocity distribution in the DF model). We therefore computed $\Delta_v$
for the
gauss model in two different ways: first, as for all the other models, by
calculating the difference between the velocity distributions at the initial
time and at time t (the curve ``gauss''), and second, by calculating the
difference between the velocity distribution at time $t$ and the equilibrium
velocity distribution, using the initial velocity distribution for the DF
model (the curve ``gauss1'').}
\label{fig_plum_drdv}
\end{figure}

Figure~\ref{fig_plum_drdv} shows the time dependences of $\Delta_r$ and
$\Delta_v$ for
our four models. It is obvious that the values of $\Delta_r$ and 
$\Delta_v$ for models DF, li,
and epol are close to their natural noise levels. At the same time, the
$\Delta_r$
value for the gauss model appreciably exceeds the noise level; $\Delta_v$ also
appreciably exceeds the noise level, although we note that the gauss model
adjusts to the equilibrium state in about 15-20 time units (which
approximately correspond to the crossing time for the entire system), and
this regime has virtually the same profile in the equilibrium state and at
the initial time.

We can draw the following general conclusion. The models
constructed using the iterative method (li and epol) are essentially in
equilibrium. They conserve their parameters as well as the model constructed
using an equilibrium distribution function (model DF).
The iterative models behave much better than the model constructed in the
approximation of a Gaussian velocity distribution (model gauss).

Our models
exhibit deviations from equilibrium at the very edge of the system, at $r > 3$
(as we would expect; Fig.~\ref{fig_plum_conv}).
On the whole, these models are very close to
equilibrium. We emphasize again that we used cold models, which are very far
from equilibrium, as the initial models for our iterations. However, through
the iterations, these models came to virtually equilibrium states. Thus, the
iterative approach has fully validated itself. In the next section, we apply
this approach to more complex systems.

\subsection{Equilibrium models of stellar disks}

\subsubsection{Implementation of the iterative method to construct an
equilibrium model for a stellar disk}
\label{s_new_methodic}
Let us apply the iterative method
described above to construct an equilibrium ($N$-body) model for a stellar
disk with a density distribution $\rho_{\rm disk}(R,z)$ embedded in an
external potential $\Phi_{\rm ext}(R,z)$. We would expect a family of
equilibrium disks to exist for a given 
$\rho_{\rm disk}$ and $\Phi_{\rm ext}$.
This must be at least a one-parameter family characterized by
the fraction of kinetic energy contained in random motions. This is where
the construction of an equilibrium stellar disk differs from the
construction of an equilibrium isotropic Plummer sphere -- we must
construct an entire family and not just one model.

The main difficulty in
the iterative method is how to transfer the velocity distribution function
(step 3 -- see the beginning of Section~\ref{s_iterative_method}).
This is done using the moments
of the distribution function and certain assumptions about the form of the
velocity distribution function (here, we have some similarity to the
Hernquist method). We did this as follows.

We took the disk model from which
we planned to ``copy'' the velocity distribution function (a slightly evolved
model from the previous iteration step). We then subdivided the model into
concentric cylindrical layers of infinite length along the $z$ axis. We
computed in each layer the four moments of the distribution function
$\bar v_{\varphi}$, $\sigma_R$, $\sigma_{\varphi}$, and $\sigma_z$,
and used these moments to specify the velocities in the model for
the next iteration step. We assumed that the velocity distribution has a
Schwartzschild form except for one feature: we removed all particles capable
of escaping beyond the disk boundary.

Let us explain this last condition. If
we have specified an axisymmetric potential $\Phi_{\rm tot}$ 
(in our case $\Phi_{\rm tot}=\Phi_{\rm disk} + \Phi_{\rm ext}$)
and a particle is specified to have the coordinates $R$, $z$ and velocities
$v_{R}$, $v_{\varphi}$, $v_{z}$, this particle can reach the radius
$R_{\rm max}$ (the disk boundary) if and only if the following
inequality \cite{binney-1987}(p. 117) is satisfied:
\be
0.5(v_{R}^2 + v_z^2 + v_{\varphi}^2) + \Phi_{\rm tot}(R,z) \ge
\Phi_{\rm tot}(R_{\rm max},0) + \frac{v_{\varphi}^2 R^2}{2 R_{\rm max}^2} \, .
\ee

We then searched for the parameters of this so ``truncated'' Schwartzschild
distribution ($\bar v_{\varphi}^\prime$, $\sigma_R^\prime$,
$\sigma_{\varphi}^\prime$, $\sigma_z^\prime$), that make its moments equal
to the specified 
values ($\bar v_{\varphi}$, $\sigma_R$,
$\sigma_{\varphi}$, $\sigma_z$). In some cases, this is impossible
in the peripheral regions of the galaxy. We then used in these regions an
``energy-truncated'' Schwartzschild distribution instead of the
radius-truncated distribution, i.e., we used the Schwartzschild distribution
without particles with positive total energies (particles capable of
escaping to infinity).

This yields a velocity distribution function in each
cylindrical layer, thereby solving the problem of transferring the velocity
distribution function. Another technical point is the following. In each
cylindrical layer, we have a distribution function with the parameters
$\bar v_{\varphi}^\prime$,
$\sigma_R^\prime$, $\sigma_{\varphi}^\prime$, $\sigma_z^\prime$.
We referenced the value of each of these parameters to
the value at the midpoint of the cylindrical layer and interpolated them
using piecewise-linear functions to obtain a velocity distribution function
that is continuous in radius.

We adopted the following important simplifying
assumptions during the transfer of the velocity distribution function (that
may fail in real stellar disks of spiral galaxies):

\begin{itemize}
\item
the disk is isothermal in the $z$ direction;
\item 
by adopting the Schwartzshild velocity
distribution, we implicitly assume that $\overline{v_R v_z}=0$.
\end{itemize}
 
We wish to emphasize two
important points. First, the particular implementation of the iterative
method used does not matter -- it is the idea behind the method that is
important. Second, the main test of any method for constructing an
equilibrium $N$-body model should, naturally, be experimentally verifying the
extent to which the model is close to equilibrium.
 
We used the following
model versions as initial models for the iterations.

\begin{itemize}
\item 
Models constructed
using the Hernquist technique (see Section~\ref{s_moments}).
\item
Cold models in which the
particles move in strictly circular orbits. This model is theoretically in
equilibrium, but is highly unstable, even on time scales of the order of the
iteration time. Therefore, this model becomes slightly heated in the
iteration process.
\item
A ``shifted'' model. Let us assume that we have an
equilibrium disk model constructed using the iterative method. In our
technique, this model is determined by the four moments of the velocity
distribution function in each cylindrical layer ($\bar v_{\varphi}$,
$\sigma_R$, $\sigma_{\varphi}$, $\sigma_z$). We construct a
new model with $\bar v_{\varphi}^\prime=c \bar v_{\varphi}$,
$\sigma_R^\prime= \frac{1}{c} \sigma_R$,
$\sigma_{\varphi}^\prime= \frac{1}{c} \sigma_{\varphi}$,
$\sigma_z^\prime=\sigma_z$, where $c$ is the shift
coefficient, which is close to unity, the
primed moments refer to the shifted model, and the unprimed moments refer to
the old, unshifted model.
\item
Other versions. For example, a model
constructed for a different external potential or for a disk with other
parameters.
\end{itemize}

\begin{figure}
\centerline{\psfig{file=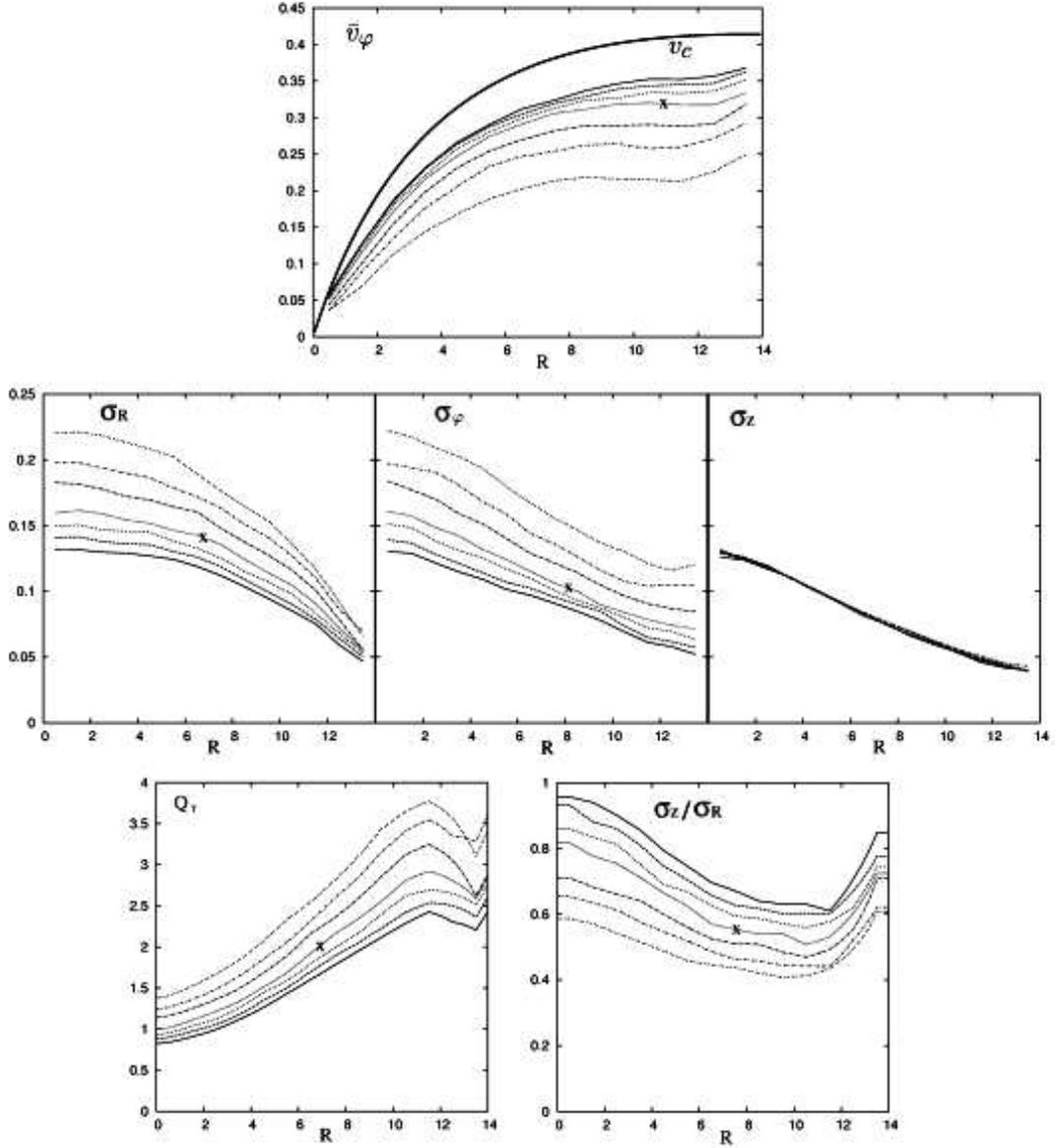,width=16cm}}
\caption{
First family of models constructed using the iterative method.
Dependences on cylindrical radius $R$ are shown for four moments of the
velocity distribution ($\bar v_{\varphi}$, $\sigma_R$,
$\sigma_{\varphi}$, and $\sigma_z$; two top rows), and the Toomre
parameter $Q_{\rm T}$ and ratio $\sigma_z/\sigma_R$ (bottom row). 
The cross indicates the model whose
equilibrium test is shown in Fig.~\ref{fig_new_model1_stability}.
The various curves in the plots
correspond to various models in this family.}
\label{fig_new_model1}
\end{figure}

\begin{figure}
\centerline{\psfig{file=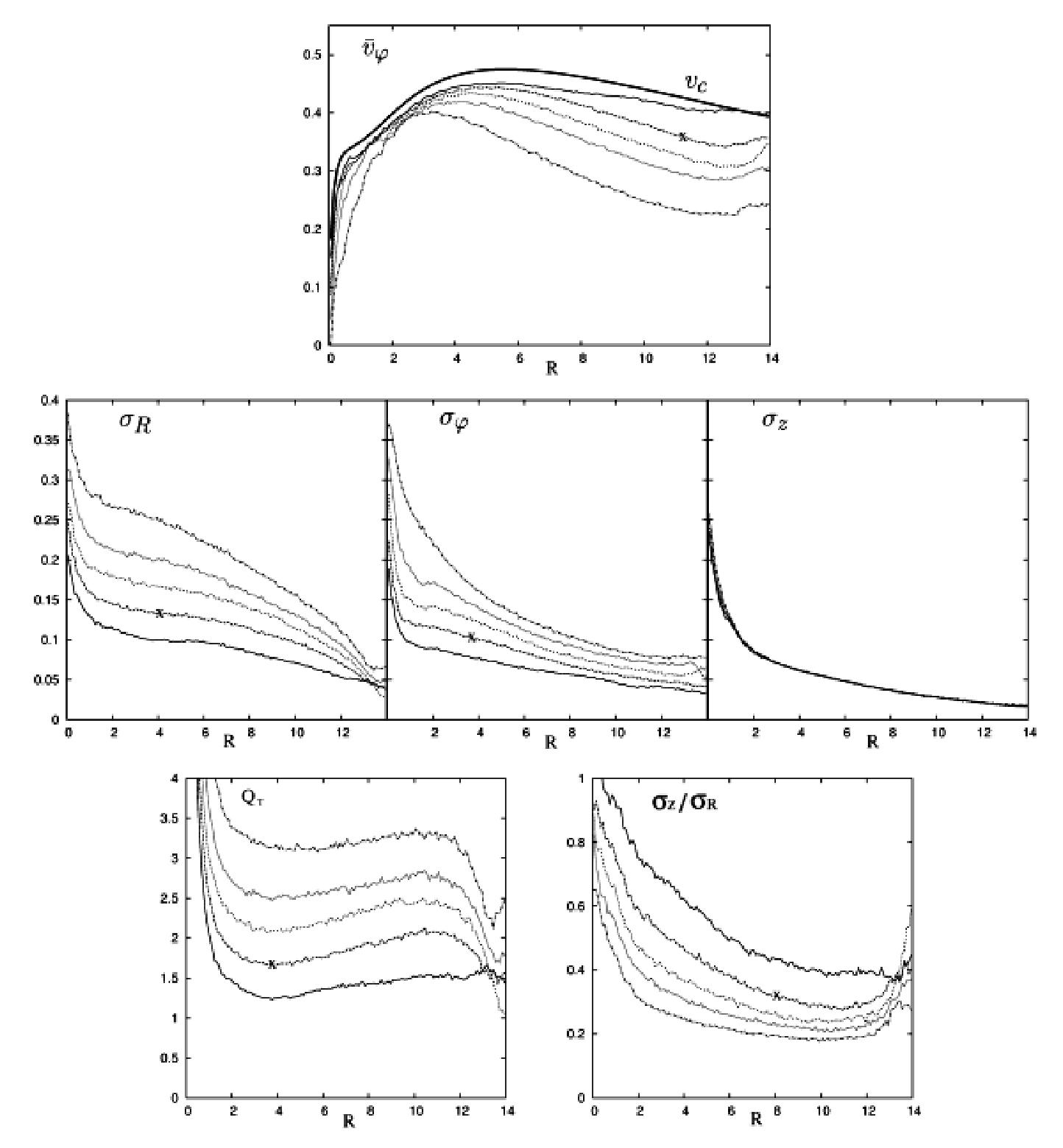,width=16cm}}
\caption{Same as Fig.~\ref{fig_new_model1} 
for the second family of models. The cross indicates the model whose
equilibrium test is shown in Fig.~\ref{fig_new_model2_stability}.}
\label{fig_new_model2}
\end{figure}

The iterations for all these versions converged and yielded
various models. However, as expected, all these models form a one-parameter
family (Figs.~\ref{fig_new_model1}~and~\ref{fig_new_model2}; 
see below for a more detailed description of these
models).

Before starting our discussion of the properties of the iterative
models, let us point out a few technical details. We must usually construct
either several models in a family with the same $\rho_{\rm disk}$ 
and  $\Phi_{\rm ext}$ or some
particular model of this family. One example might be a model with a given
kinetic energy fraction contained in the random motions of stars (in this
case,
we can fix the Toomre parameter at some radius). The convergence of the
iterations is faster the closer the initial disk is to equilibrium.
Therefore, to save CPU time, it is advisable to use the following ``tricks.''
We can initially perform the iterations with lower accuracy (fewer particles
and a coarser dividing of the model into cylindrical layers during the
transfer of the particle velocity distribution), and then carry out several
iterations with higher accuracy. If we have already constructed one
equilibrium model from a family, all the other models of this family can
conveniently be constructed using shifted models (described above, where we
enumerated possible initial models for the iterations). Such shifted models
turn out to be much closer to equilibrium than, for example, models
constructed using the Hernquist technique, and the iterations converge
appreciably faster.

To accelerate the convergence of the iterations and
construct models with a fixed fraction of their kinetic energy contained in
random motions, we can fix this energy fraction at each iteration (i.e., fix
the degree of heating of the model). Instead of fixing the amount of energy
contained in random motions, we can fix the amount of energy
contained in regular motions. We fixed this latter quantity via the total
angular momentum with respect to the $z$ axis:
\be
L_z = \sum_{i=1}^N m_i v_{\varphi i} R_i \, ,
\ee
where $m_i$, $v_{\varphi i}$, and $R_i$ are the mass, azimuthal velocity, and cylindrical
radius of the $i$th particle, respectively.

 We fixed $L_z$ and, at each iteration
step, having constructed the new model (with the same velocity distribution
as in the slightly evolved model of the previous iteration step), we
adjusted the azimuthal velocities of the particles to make the total angular
momentum of the system exactly equal to the given $L_z$. We did this as
follows. Let $L_z$ be the given angular momentum and $L_z^\prime$,
the current angular
momentum. We specified the new azimuthal velocities of the particles using
the relations
\be
v_{\varphi i} = v_{\varphi i}^\prime + \frac{(L_z - L_z^\prime)}{R_i m_i} 
\frac{w_i}{\sum_{j=1}^{N}w_j} \, ,
\ee
where $v_{\varphi i}^\prime$ is the current azimuthal velocity of the $i$th particle,
$v_{\varphi i}$ the
adjusted azimuthal velocity of this particle,
and $w_i$ a weight that
determines how we add the angular momentum to the particle. In our
experiments, we set $w_i=R_i m_i$.

Iterations with fixed $L_z$ converged to models
of the same family as the iterations without fixed $L_z$. However, the
iterations with fixed $L_z$ converged much
faster. Fixing $L_z$ is helpful, since it results in the convergence of the
iterations to a strictly defined model from a family of iterative models
with given $\rho_{\rm disk}$ and $\Phi_{\rm ext}$, irrespective of the initial model.

\subsubsection{Properties of the iterative disk models}

Let us consider as examples two
families of models constructed using the iterative method.

Figure~\ref{fig_new_model1} shows
the first family. We adopted disk model (\ref{eq_exp_disk_dens}) 
with $h=3.5$, $z_0=1$, $M_{\rm disk}=1$, and $R_{\rm max}=14$,
and modeled the external potential using the Plummer sphere
(\ref{eq_plum_phi}) with $a_{\rm pl} = 15$ and $M_{\rm pl}=4$.
For these parameters, the fractional mass of
the spherical subsystem within four exponential radii is equal to about $1.3$
(i.e., $M_{\rm sph}(4h)/M_{\rm disk}(4h) \approx 1.3$). 
Note that the parameters of the disk and
external potential are the same as in the model constructed using the
Hernquist technique (Fig.~\ref{fig_old_model}), which was
used as an example of a not-quite-equilibrium model. The time for a single
iteration, $t_i=15$, corresponds to approximately one-tenth of a disk
rotation at a distance of two exponential disk scales. The thickness of the
cylindrical layers into which we subdivided the model when transfering the
velocity distribution function is $d_R=1$. We used $N=10^5$ particles.

Figure~\ref{fig_new_model2}
shows the second family. It was produced using an initially thinner disk
model (\ref{eq_exp_disk_dens}) with 
$h=3.5$, $z_0=0.3$, $M_{\rm disk}=1$, and $R_{\rm max}=14$, modeling the
external potential using the Plummer sphere (\ref{eq_plum_phi}) 
with  $a_{\rm pl} = 3.5$, $M_{\rm pl}=0.88$ and the potential of a Hernquist 
\cite{hernquist-359-1990} sphere
\be
\label{eq_hernq_phi}
\Phi(r) = -\frac{G M_{\rm hr}}{r + a_{\rm hr}}\, ,
\ee
with $a_{\rm hr} = 0.5$ and $M_{\rm hr}=0.2$. In this case, the
Plummer and Hernquist spheres play the role of a dark halo and bulge,
respectively. The total relative mass of the spherical subsystem within four
exponential disk parameters is approximately unity 
($M_{\rm sph}(4h)/M_{\rm disk}(4h) \approx 1$).
The duration of a single iteration is $t_i=15$; the thickness of the
cylindrical layers into which we subdivided the model for the transfer of
the velocity distribution function is $d_R=0.1$ (i.e., a less coarse
dividing than in the first case), and the number of particles is 
$N=5 \cdot 10^5$.

The models constructed using the iterative method have the following
properties.

\begin{figure}
\centerline{\psfig{file=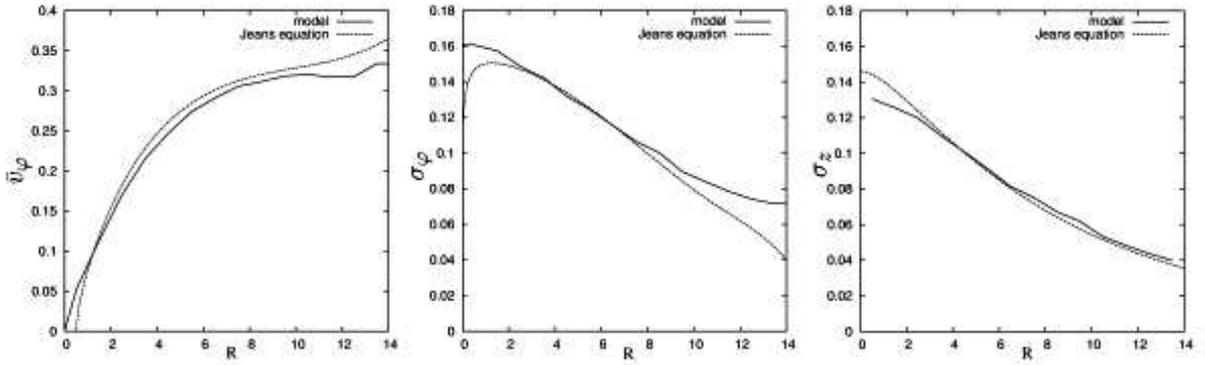,width=16cm}}
\caption{
Comparison of the moments of the velocity distribution function
for the iterative model of the first family with the solutions of the Jeans
equations (\ref{eq_jeans_1}).
In the left-hand plot, the solid curve (model) shows the mean
azimuthal velocity for the iterative model, and the dashed curve (Jeans
equation) shows the same quantity computed proceeding from the Jeans
equation [the first equation of (\ref{eq_jeans_1})],
where we adopted $\sigma_R$ and $\sigma_{\varphi}$  from the
iterative model. In the middle plot, the solid curve shows the azimuthal
velocity dispersion for the iterative model, and the dashed curve the same
quantity computed using the Jeans equations [the second equation of
(\ref{eq_jeans_1})],
where we adopted $\bar v_{\varphi}$ and $\sigma_R$ from the iterative model. 
In the right-hand plot,
the solid curve shows the dispersion of the stellar velocities in the
vertical direction, and the dashed curve the same quantity computed using
the Jeans equations [the third equation of (\ref{eq_jeans_1})]. We used a model of the
family presented in Fig.~\ref{fig_new_model1} (marked by a cross).}
\label{fig_new_model1_jeans}
\end{figure}

\begin{figure}
\centerline{\psfig{file=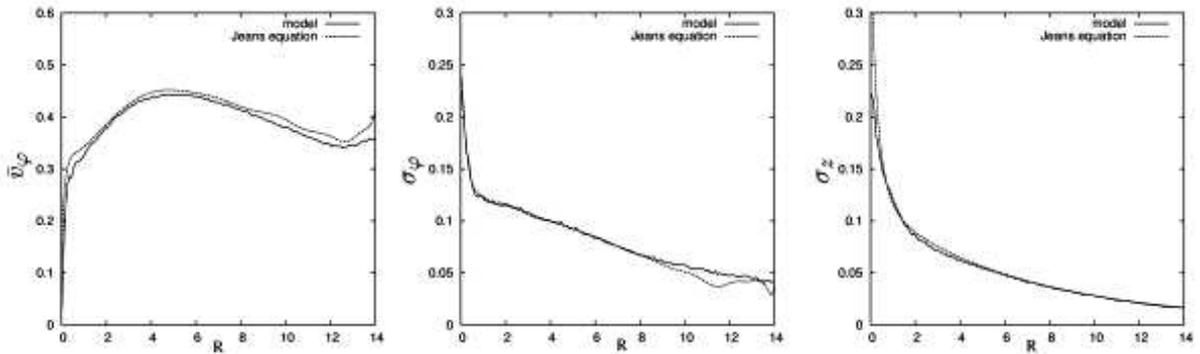,width=16cm}}
\caption{
Same as Fig.~\ref{fig_new_model1_jeans} 
for the iterative model of the second family. We
used a model of the family shown in Fig.~\ref{fig_new_model2} (marked by a cross).
}
\label{fig_new_model2_jeans}
\end{figure}

\begin{figure}
\centerline{\psfig{file=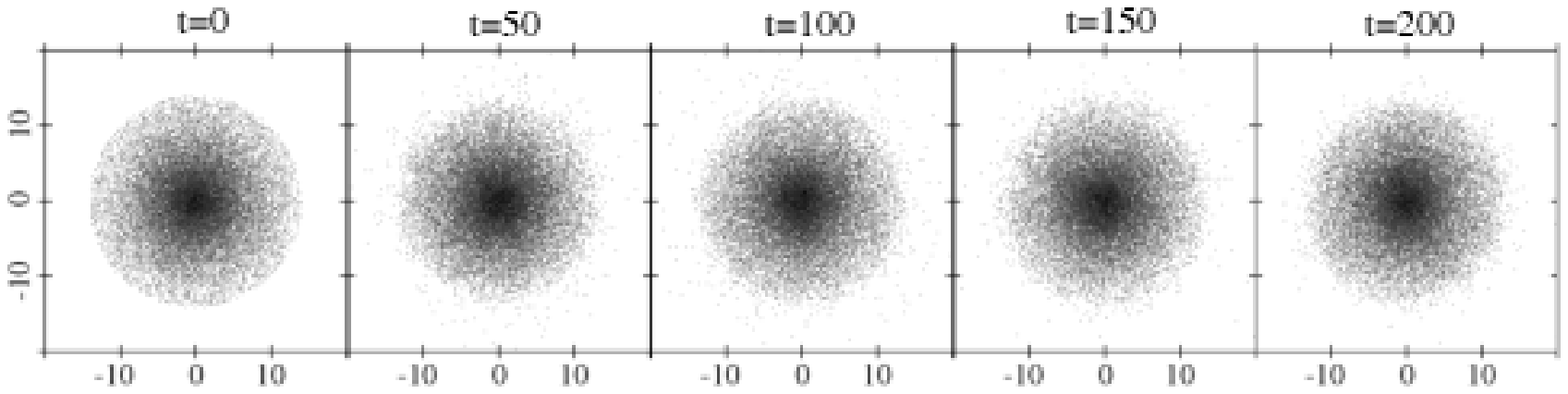,width=16cm}}
\centerline{\psfig{file=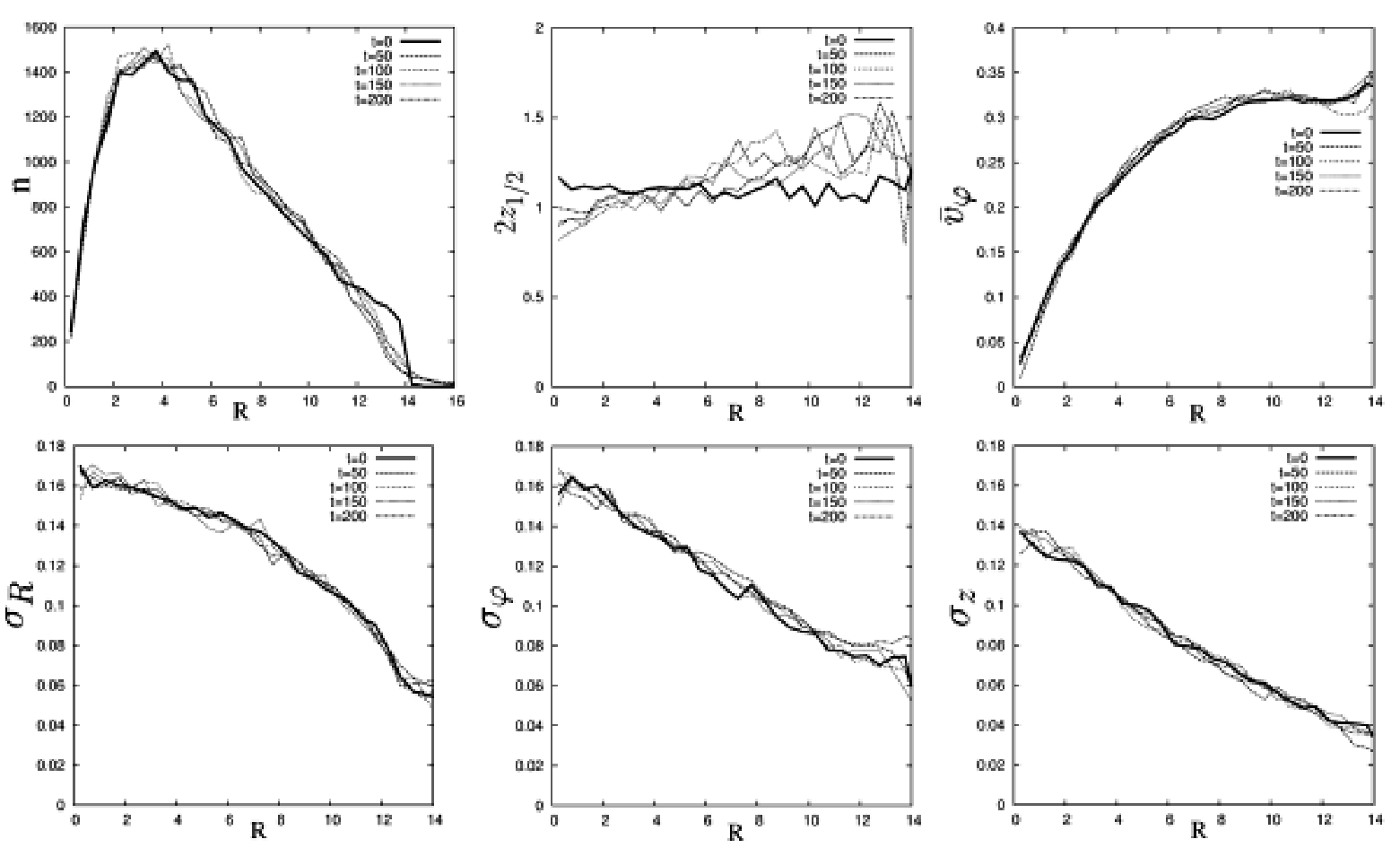,width=16cm}}
\caption{ Initial stages of the evolution of an iterative model in the
first family. A model of the family shown in Fig.~\ref{fig_new_model1}
(marked by a cross) is used. The top snapshots show a face-on view for
several times, with the shades of gray corresponding to the logarithm of the
particle number per pixel. The
plots in the central and bottom parts of the figure show the dependences of
various disk parameters on the cylindrical radius $R$ for various times: the
number of particles in concentric cylindrical layers $n$, twice the median of
$|z|$, $2 z_{1/2}$ (which characterizes the disk thickness and is close to
$z_0$ for 
the distribution (\ref{eq_exp_disk_dens})), and the four moments of the
velocity distribution 
function $\bar v_{\varphi}$, $\sigma_R$, $\sigma_{\varphi}$, and $\sigma_z$.
$N = 25\,000$, $\epsilon=0.02$, and $dt=0.01$. The disk 
thickness can be seen to have changed slightly. This is probably due to the
assumption that the $z$-velocity distribution function is isothermal.}
\label{fig_new_model1_stability}
\end{figure}

\begin{figure}
\centerline{\psfig{file=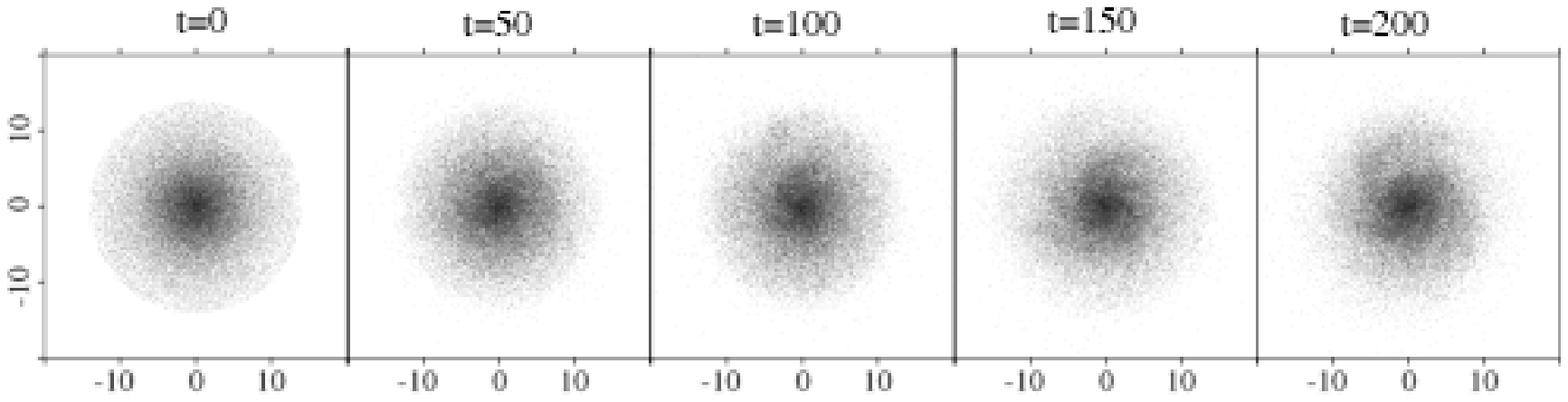,width=16cm}}
\centerline{\psfig{file=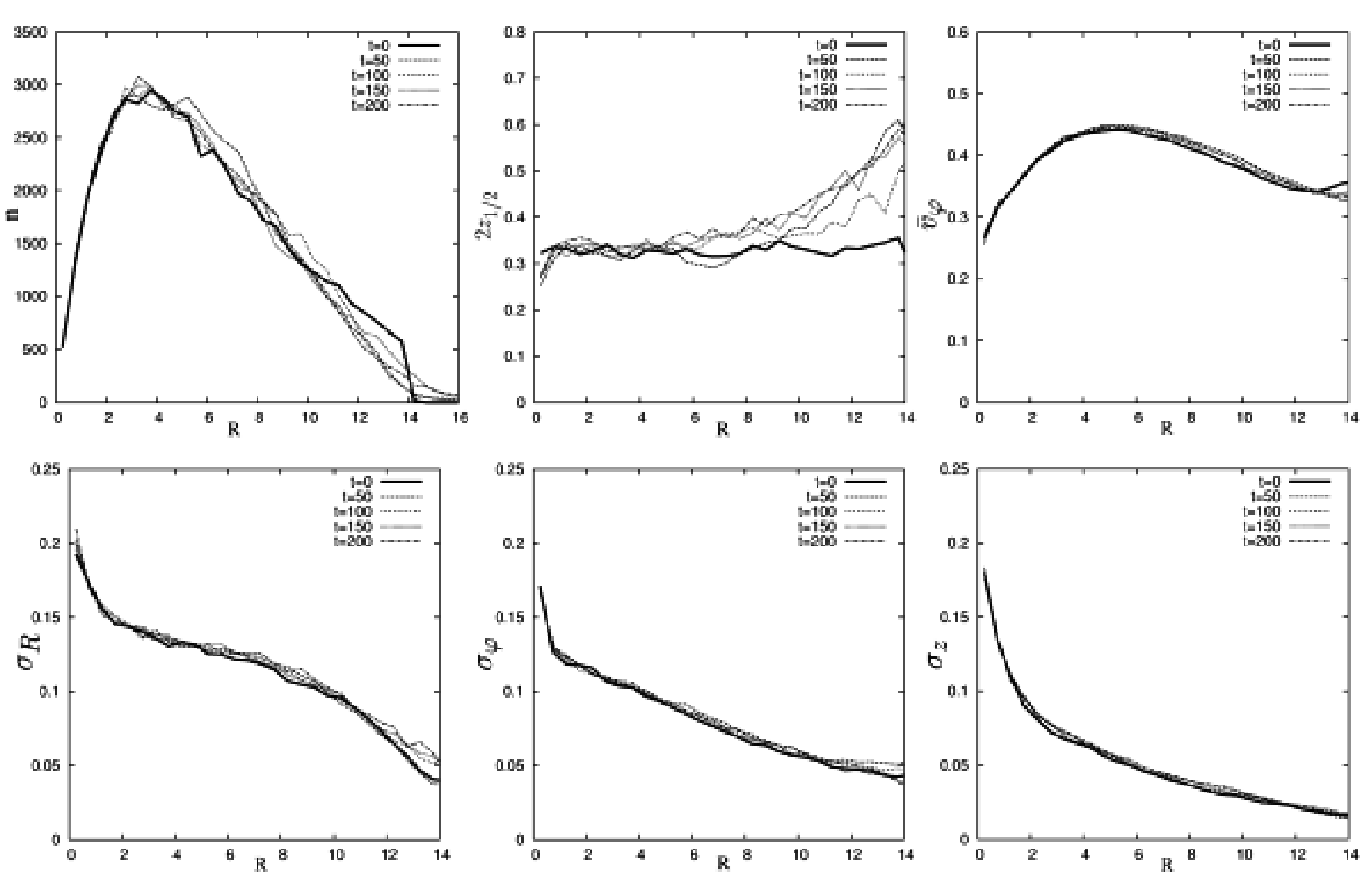,width=16cm}}
\caption{
Same as Fig.~\ref{fig_new_model1_stability} for an iterative model of the second family 
($N = 50000$, $\epsilon=0.02$, $dt=0.01$).
A model of the family shown in Fig.~\ref{fig_new_model2} (marked by a cross)
is used.}
\label{fig_new_model2_stability}
\end{figure}

\begin{enumerate}
\item
The models prove to be close to equilibrium. They preserve
both their structural and dynamical parameters well during the initial
stages of their evolution
(Figs.~\ref{fig_new_model1_stability}~and~\ref{fig_new_model2_stability}). 
It goes without saying that
different values of $\rho_{\rm disk}$ and $\Phi_{\rm ext}$ yield
models with different degrees of closeness to equilibrium. For example, the
models of a thin stellar disk without an external potential (with the
density distribution
(\ref{eq_exp_disk_dens} and $h=3.5$, $z_0=0.3$, $M_d=1$, $R_{\rm max}=14$)
are appreciably nonequilibrium. However, in any case, the resulting models
are substantially closer to equilibrium than those constructed using the
Hernquist technique (see Section~\ref{s_moments}). We consider possible reasons for
deviations of the iterative models from equilibrium below.

\item
The first
four moments of the velocity distribution function fairly accurately obey
the equilibrium Jeans equations (\ref{eq_jeans_1}) 
(Figs.~\ref{fig_new_model1_jeans}~and~\ref{fig_new_model2_jeans}).
This means that, if we
replaced the assumed dependence $\sigma_R(R)$ in the Hernquist method 
(see Section~\ref{s_moments})
with the dependence obtained from the iterations, the resulting models will
be very similar to the iterative models,
i.e., they essentially be in equilibrium. In other words, given a more
correct assumption about the dependence $\sigma_R(R)$,
the Hernquist technique would
allow the construction of close-to-equilibrium models.
\item
All models in
some family of iterative models have the same vertical-velocity dispersion
profile $\sigma_z(R)$ (Figs.~\ref{fig_new_model1}~and~\ref{fig_new_model2}).
This means that the equilibrium condition in the
vertical direction is completely decoupled from the disk equilibrium
condition in the plane. This was expected, since, of all the moments of the
velocity distribution function, only $\sigma_z$ appears in the Jeans equation
describing equilibrium in the vertical direction [the last equation of
(\ref{eq_jeans_1})]. 

\item
The profile of the radial stellar-velocity dispersion, $\sigma_R$, 
differs strongly from the commonly adopted exponential profile
(Figs.~\ref{fig_new_model1}~and~\ref{fig_new_model2}).
\end{enumerate}

\subsubsection{Specific features of the iterative disk models. Further
development of the technique}

Let us now turn to some specific features of our iterative
models.

\begin{enumerate}
\item 
Our stellar disk model (\ref{eq_exp_disk_dens}) is not a very good choice. The
density profile in this model has a sharp cutoff. Naturally, such a
structure cannot be in equilibrium, and the disk must have a smooth cutoff.
We can see in
Figs.~\ref{fig_new_model1_stability}~and~\ref{fig_new_model2_stability}
(the left-hand plot in the middle row) how this
sharp cutoff is disrupted. We conclude that we must choose an initial
stellar-disk model with a smooth density decrease at $R$ values close to
$R_{\rm max}$.

\item
As is clear from
Figs.~\ref{fig_new_model1_stability}~and~\ref{fig_new_model2_stability}
(middle plot in the middle row), the
thickness of the stellar disk varies somewhat during the initial stages of
evolution. These variations can be seen in almost all the iterative
models, and are more pronounced in models that are hotter in the disk plane
and less pronounced in models that are cooler in the disk plane (here,
hotter means that a higher fraction of the kinetic energy is contained in
random motions). This nonequilibrium state of the iterative models is likely
due to the assumptions about the velocity distribution function adopted when
implementing our method, in particular our assumption that the velocity
distribution function does not depend on $z$ (the model is isothermal in the
$z$ direction). In addition, the assumption that $\overline{v_R v_z}=0$ may
not be valid in the central regions. Future analyses
should apply the iterative method without these assumptions, first and
foremost, without the isothermal assumption, especially since this method is
easy to generalize in this way. In the technique described here, we
subdivided the model into cylindrical layers when transfering the velocity
distribution function from the previous iteration step. Each of these
cylindrical layers can be subdivided into layers along the $z$ axis, and the
moments of the velocity distribution function can be computed
separately in each layer (i.e., the model can be subdivided in both the $R$
and $z$ planes). Note, however, that such
a modification of the iterative method requires a far greater number of
particles than we have used so far.

\item
The iterative method cannot yield an
equilibrium model that is unstable on time scales shorter than the time for
one iteration. For example, the method cannot produce a very cold model. As
is evident in Figs.~\ref{fig_new_model1}~and~\ref{fig_new_model2}, 
the family of iterative models is bounded on
the side of cold models. It is not possible to construct a very cold model,
even when $L_z$ is fixed -- the iterations still fail to converge.
\end{enumerate}

\subsubsection{Assumption of uniqueness of the family of models. 
New applications for iterative models}

Thus, the technique described above can be used to
construct a one-parameter family of close-to-equilibrium models for a given
$\rho_{\rm disk}$ and $\Phi_{\rm ext}$. This family is parametrized by the
fraction of kinetic energy 
contained in random motions (e.g., this parameter
can have the form of the total angular momentum of the disk with respect to
the $z$ axis, $L_z$). The question naturally arises in this case of whether there
are other equilibrium models besides those in the family of iterative disk
models? We can only postulate that there can be only one equilibrium disk
model. Such a hypothesis can be formulated follows. There can be at most one
equilibrium model (one equilibrium distribution function) for a given
$\rho_{\rm disk}(R,z)$, $\Phi_{\rm ext}(R,z)$ with a fixed fraction of
kinetic energy contained in random motions (e.g., for fixed $L_z$).

We can neither prove nor disprove this
statement, and can only assume that it is true. This
assumption expands the
scope of application of the iterative models. We can compare iterative
models to observational data in order to derive constraints on unobserved
parameters of galaxies. For example, our statement that
the dependence of $\sigma_R$ on $R$ for the iterative models is far from
exponential 
requires additional explanations. More precisely, this dependence is far
from exponential for the $\rho_{\rm disk}$ and $\Phi_{\rm ext}$ considered.
It is not ruled out that 
an external potential can be selected so as to make the dependence of
$\sigma_R$ on $R$
exponential \cite{widrow-838-2005} (see also Section~\ref{s_sigma_R_z},
where we discuss the observational data 
for this dependence). When available, GAIA data on the velocity field in our
Galaxy can be used to try to constrain the mass distribution in the dark
halo of the Galaxy via iterative models (by selecting an external potential
that makes the equilibrium stellar disk embedded in this potential have the
observed velocity field).

\section{Conclusions}

We have proposed here a new
iterative approach to constructing equilibrium $N$-body
models with a given
density distribution. The main idea of this approach is the following. At
the first stage, a model is constructed using some approximate method. The
model is then allowed to adjust to the equilibrium state while its density
distribution is kept fixed, as well as the required parameters of the
velocity distribution, if necessary.

We used our iterative approach to
construct two types of models. The first type have the form of isotropic and
spherically symmetric systems. We used the iterative approach to construct
such systems in order to test whether the approach indeed enables the
construction of close-to-equilibrium models. The second type of models were
axisymmetric stellar disks in an external potential. In both cases, the
numerical models constructed using the iterative approach were very close to
equilibrium. The iterative models for the stellar disks were much closer to
equilibrium than models constructed based on the Jeans equations (see
Section~\ref{s_moments}).

We also show that the hypothesis that the radial velocity
dispersion is proportional to the vertical velocity dispersion in spiral
galaxies ($\sigma_R \propto \sigma_z$) may be incorrect. First, this
assumption lacks clear 
observational evidence to support it. Second, it is not true for the
iterative models of stellar disks that we have constructed. Finally, the
assumption that $\sigma_R \propto \sigma_z$ is the main reason why the
stellar disks constructed 
using the technique of Hernquist \cite{hernquist-389-1993} are fairly far
from equilibrium. To 
prove this last statement, recall that, since the iterative models of
stellar disks obey the Jeans equations, if we replace in Hernquist's
technique the relation $\sigma_R \propto \sigma_z$ with the dependence
$\sigma_R(R) = \sigma_R^i(R)$, where $\sigma_R^i(R)$ is
the radial dispersion
profile for an iterative model, the resulting models should be as close to
equilibrium as the iterative models.

The proposed method has a wide range of applications, and can be used to
construct models with various geometries. For example, it can be used to
construct spherically symmetric models with anisotropic stellar motions, as
well as self-consistent multicomponent models of disk galaxies. Unlike the
kinetic approach based on the distribution function, our method provides
direct input data for $N$-body simulations.

Furthermore, the proposed method
has purely astrophysical applications. For example, bringing the velocity
field obtained in equilibrium iterative models into agreement with the
velocity field in our Galaxy can enable determination of the parameters of
the external potential, and, consequently, the dark halo.

\vspace{0.5cm}
\noindent
{\large Acknowledgment}\\
\noindent

This work was supported by the Russian Foundation for Basic Research
(project codes 03-12-17152 and 06-02-16459) and the Program of Support for
Leading Scientific Schools of the Russian Federation (grant NSh-8542.2006.2).


\begin{flushright}
Translated by A. Dambis
\end{flushright}

\end{document}